\documentclass[aps,floats,twocolumn,showpacs]{revtex4-1}
\usepackage{graphicx}
\usepackage{amsmath}
\usepackage{amssymb}
 \def\Dl{{\scriptstyle \Delta}}

\begin{document}
 \title{Transition regime of the one-dimensional two-stream instability}
 \author{K.V. Lotov and I.V. Timofeev}
 \affiliation{Budker Institute of Nuclear Physics, 630090, Novosibirsk, Russia \\
 Novosibirsk State University, 630090, Novosibirsk, Russia}
 \date{\today}
 \begin{abstract}
The transition between kinetic and hydrodynamic regimes of the one-dimensional
two-stream instability is numerically analyzed, and the correction
coefficients to the well-known textbook formulae are calculated. The approximate expressions are shown to overestimate the growth rate several times in a wide parameter area.
 \end{abstract}
 \maketitle

An electron beam propagating through a dense plasma is unstable against a
longitudinal density modulation. This is a basic plasma instability known as
the two-stream instability; it is described in many textbooks (see, e.g., Refs.
\cite{Briggs,Krall,Mikh,AAIvanov}). Depending on the velocity spread of the
beam $\Dl v$ and on the wavenumber $k$ and growth rate $\gamma$ of the
unstable mode, two regimes of the instability are distinguished: hydrodinamic
($k \Dl v \ll \gamma$) and kinetic ($k \Dl v \gg \gamma$). In both regimes,
the dependence $\gamma (k)$, as well as scalings for
the maximum growth rate, can be easily found. Here we study the two-stream
instability in the transition regime ($k \Dl v \sim \gamma$), for which there
are no well-known scalings.

The interest to this classical problem is renewed by recent progress in plasma
heating by powerful electron beams \cite{OS2006-106,PPR31-462,JETPL77-358}.
There are some evidences that in these experiments the level of resonant
Langmuir waves excited by the beam could be determined by beam trapping
effects. Profiles of the energy release along the plasma column calculated
under this assumption are in a surprisingly good quantitative agreement with
experimental observations \cite{Pop13-062312}. In turn, the level at which the
wave energy saturates due to beam nonlinearity is very sensitive to the
instability growth rate. In the one-dimensional nonrelativistic case
\cite{Galeev&Sagdeev} this level scales as $\gamma^4$. Thus, for a detailed
study of beam relaxation, the growth rate of the instability needs to be known
with a good precision in a wide area of beam parameters.

We consider the simplest one-dimensional model: a non-relativistic electron
beam of the density $n_b$ and the velocity distribution
\begin{equation}\label{e1}
  f(v)=\frac{1}{\Dl v \sqrt{\pi}}\exp \left(
  -\frac{(v-v_0)^2}{\Dl v^2} \right)
\end{equation}
propagates through a cold uniform plasma of the density $n_0$. This model may
be too basic for description of real physical systems, where the growth of
obliquely propagating waves, the final width of the beam, or the presence of
an external magnetic field usually complicate the picture of the instability.
However, the simplicity of the model allows us to present the main features of
the transition regime in a visually graspable form.  The model also can be
used for testing kinetic numerical codes, for which the operation in a safely
kinetic regime may be too time consuming because of the required low beam
densities and low growth rates.

A similar problem was earlier considered in papers
\cite{PF11-1754,JGR94-2429}, but these studies were mainly concentrated on
changes in topology of the dispersion curves. Here we focus our attention on
comparison of the exact solution and its standard approximations.

Following the standard technique \cite{Shafr}, we can obtain the dispersion
relation for fast longitudinal waves and rewrite it in the form
\begin{equation}\label{e2}
  \tilde k^2 = \frac{1}{(1+\xi \Dl \tilde v)^2}
  - \frac{2 \tilde n_b}{\Dl \tilde v^2} \bigl(1 - Z(\xi)\bigr),
\end{equation}
where
\begin{equation}\label{e3}
  Z(\xi) = 2 \xi e^{-\xi^2} \int_0^\xi e^{x^2} dx
  - \imath \sqrt{\pi} \xi e^{-\xi^2}
\end{equation}
is the plasma dispersion function,
\begin{equation}\label{e4}
  \xi = \frac{\omega_r - k v_0}{k \Dl v} + \imath \frac{\gamma}{k \Dl v},
\end{equation}
$\omega_r$ is the real part of the wave frequency, and $\tilde n_b = n_b/n_0$.
We use tildes to denote dimensionless quantities; velocities are measured in
units of $v_0$; and frequencies, in units of the plasma frequency $\omega_p =
\sqrt{4 \pi n_0 e^2/m}$.

For $|\xi|\gg 1$ (hydrodynamic regime), Eq.(\ref{e2}) reduces to
\begin{equation}\label{e5}
  \tilde k^2 = \frac{1}{(1+\xi \Dl \tilde v)^2}
  - \frac{\tilde n_b}{\xi^2 \Dl \tilde v^2},
\end{equation}
from which, for a real $\tilde k$, we obtain the following familiar results: the maximum growth rate corresponds to $\tilde
k=\tilde k_m \approx 1$ (or $k_m \approx \omega_p/v_0$) and
\begin{equation}\label{e6}
  \tilde{\omega}_r (\tilde{k}_m) \approx 1- \frac{\tilde n_b^{1/3}}{2^{4/3}},
  \qquad
  \tilde{\gamma} (\tilde{k}_m) \approx 0.69 \, \tilde n_b^{1/3}.
\end{equation}
The limit $|\xi|\ll 1$ is known as the kinetic regime. Here we can put
$Z(\xi)\approx - \imath \sqrt{\pi} \xi e^{-\xi^2}$ and find
\begin{equation}\label{e7}
  \tilde{\omega}_r (\tilde{k}) \approx \frac{1}{\sqrt{1+2 \tilde n_b /
  (\tilde k^2 \Dl \tilde v^2)}},
\end{equation}
\begin{equation}\label{e8}
  \tilde{\gamma} (\tilde{k}) \approx \frac{\tilde n_b \sqrt{\pi} \,
  (\tilde k-1)}{\tilde k^3 \Dl \tilde v^3} \,
  \exp\left(-\frac{(1-\tilde k)^2}{\tilde k^2  \Dl \tilde v^2}\right).
\end{equation}
For $\Dl \tilde v \ll 1$, the expression (\ref{e8}) has its maximum at
\begin{equation}\label{e9}
  \tilde{k}=\tilde{k}_m \approx 1 + \Dl \tilde v / \sqrt{2}, \qquad
  \tilde{\gamma} (\tilde{k}_m) \approx 0.76 \, \tilde n_b / \Dl \tilde v^2.
\end{equation}

\begin{figure}[t]
\centering
\includegraphics*[width=214bp]{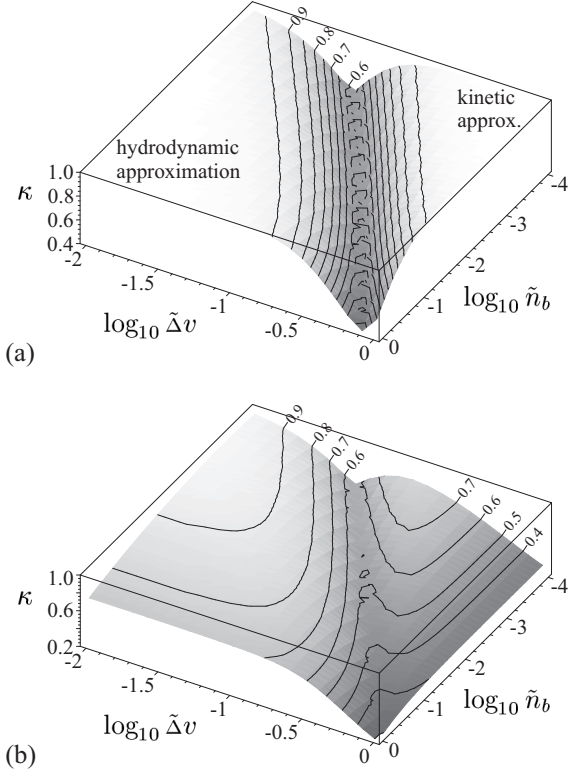}
\caption{ The correction factor to exact (a) and approximate (b) expressions
for the maximum growth rate obtained within hydrodynamic or kinetic
approximations.} \label{isurf}
\end{figure}

At arbitrary $\tilde n_b$ and $\Dl \tilde v$, we solve equation (\ref{e2})
numerically to yield the functions $\omega_r (\tilde k)$ and $\gamma (\tilde
k)$, as well as the dependence of the maximum growth rate $\gamma_m = \gamma
(\tilde k_m)$ and its location $\tilde k_m$ on the beam parameters $\tilde
n_b$ and $\Dl \tilde v$. The function $\gamma_m (\tilde n_b, \Dl \tilde v)$
itself is not very informative, so we plot in Fig.~\ref{isurf}a the correction
factor $\kappa$, the ratio of the exact $\gamma_m$ to the maximum growth rate
obtained by solving Eq.(\ref{e5}) or maximization of (\ref{e8}), whichever
gives better approximation. For comparison, in Fig.~\ref{isurf}b there is a
ratio of $\tilde{\gamma}_m$ to approximations of the maximum growth rate given
by (\ref{e6}) or (\ref{e9}), whichever is better. We see that, in a wide
parameter area, the simple formulae (\ref{e6}) and (\ref{e9}) are only
order-of-magnitude correct, while unabridged hydrodynamic and kinetic models
give reasonably good approximations within their applicability areas. It
should be particularly emphasized that the key simplifying assumptions used in
unabridged hydrodynamic or kinetic models lead to overstatement of the growth
rate which amounts to factor of two in the transition region.

\begin{figure}[t]
\centering
\includegraphics*[width=182bp]{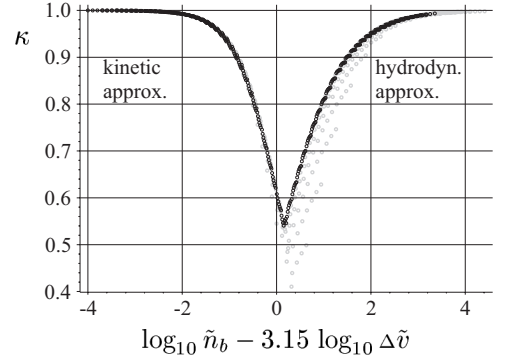}
\caption{ Dependence of the correction factor $\kappa$ on the combined beam
parameters for $1 > \Dl \tilde v > 0.01$. The black points correspond to
$10^{-2} > \tilde n_b > 10^{-4}$; the grey points, $1 > \tilde n_b > 0.01$.}
\label{idiff}
\end{figure}
Surprisingly, the ``valley'' in Fig.~\ref{isurf}a is straight and has an
invariable cross-section. Thus, with a good precision, we may assume that the
beam parameters enter the function $\kappa (\tilde n_b, \Dl \tilde v)$ only as
a combination $\tilde n_b \Dl \tilde v^{-s}$, with $s \approx 3.15$ found
empirically. The dependence of the correction factor $\kappa$ on the
difference of decimal logarithms $\log_{10} \tilde n_b - s \log_{10} \Dl
\tilde v$ is shown in Fig.~\ref{idiff}.

\begin{figure}[t]
\centering
\includegraphics*[width=199bp]{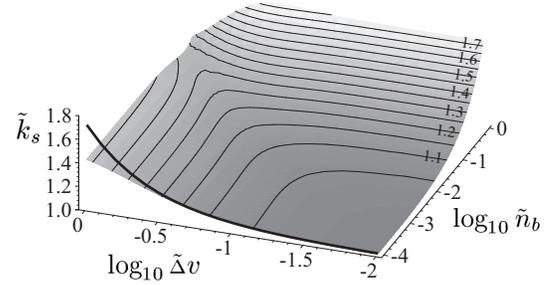}
\caption{ Wavenumber $\tilde k_m$ of the most unstable wave. The thick line
shows the scaling (\protect\ref{e9}).} \label{loc}
\end{figure}
For $\tilde n_b \ll 0.1$, the wavenumber $\tilde k_m$ of the most unstable
mode does not depend on the beam density, follows the kinetic formula
remarkably well, and, at $\Dl \tilde v \ll 1$, can be safely approximated by
(\ref{e9}) (Fig.~\ref{loc}). As the beam density approaches the plasma
density, $\tilde k_m$ tends to $\sqrt{3}$, the value found for equal
counterstreaming electron flows \cite{Mikh}.

\begin{figure}[bth]
\centering
\includegraphics*[width=231bp]{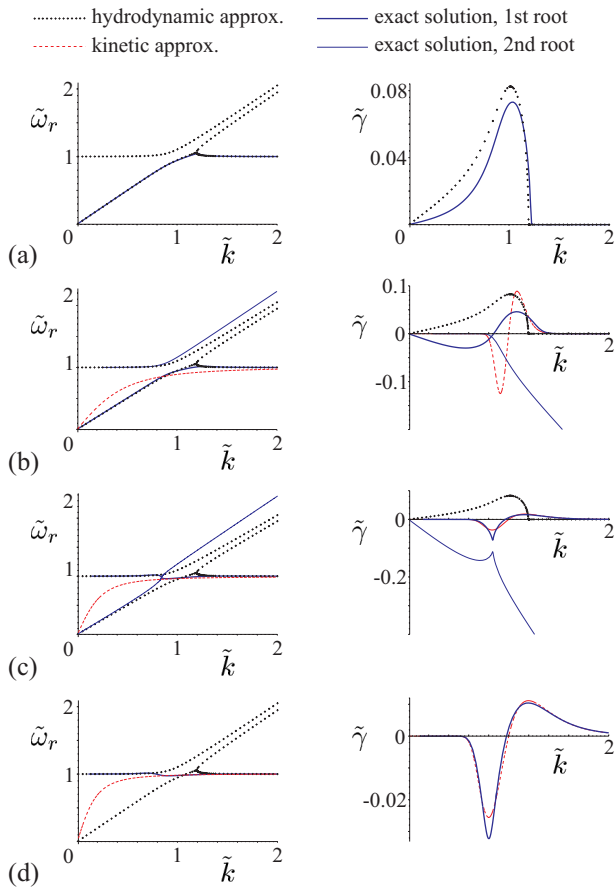}
\caption{(Color online) Evolution of dispersion curves: numerical solution to the full
dispersion relation (\protect\ref{e2}) (solid lines), numerical solution to
hydrodynamic dispersion relation (\protect\ref{e5}) (dotted lines), and
kinetic expressions (\protect\ref{e7}) and (\protect\ref{e8}) (dashed lines):
real part $\tilde \omega_r (\tilde k)$ (left column) and growth rate $\tilde
\gamma (\tilde k)$ (right column). Beam density $\tilde n_b = 0.002$, beam
velocity spread: (a)
 $\Dl \tilde v = 0.05$ (nearly hydrodynamic regime), (b) $\Dl \tilde v = 0.12$
(the greatest difference between maximum growth rates), (c) $\Dl \tilde v =
0.24$ (close to reconnection of dispersion curves), and (d) $\Dl \tilde v =
0.3$ (nearly kinetic regime). } \label{curves}
\end{figure}

It is interesting to follow evolution of exact and approximate dispersion
curves as we travel in the parameter space from the hydrodynamic regime to the
kinetic one. For this, we put $\tilde n_b = 0.002$ and change $\Dl \tilde v$
(Fig.~\ref{curves}). As $\Dl \tilde v$ increases, the growth rate decreases
(Fig.~\ref{curves}a), the instability interval shifts to greater values of
$\tilde k$ and narrows, while the real part of the frequency changes
insignificantly (Fig.~\ref{curves}b). At some value of $\Dl \tilde v$ (when
the growth rate already approaches the value given by the kinetic
approximation), a reconnection of dispersion curves takes place
(Fig.~\ref{curves}c) \cite{PF11-1754,JGR94-2429}, after which both real and
imaginary parts of the wave frequency closely follows the kinetic formulae
(\ref{e7}) and (\ref{e8}).

This work is supported by RF President's grants NSh-2749.2006.2 and
MD-4704.2007.2, RFBR grants 06-02-1657 and 08-01-00622, and Russian Ministry of Education (projects RNP 2.2.1.1.3653 and innovation educational project 456).

\end{document}